**The M³ project: 1- A global hyperspectral image-cube of the Martian surface**

Riu L., Poulet F., Carter J., Bibring J.-P., Gondet B., M. Vincendon

**Abstract**

This paper is the first paper of a series that will present the derivation of the modal mineralogy of Mars (M³ project) at a global scale from the near-infrared dataset acquired by the imaging spectrometer OMEGA (Observatoire pour la Minéralogie, l'Eau, les Glaces et l'Activité) on board ESA/Mars Express. The objective is to create and provide a global 3-D image-cube of Mars at 32px/° covering most of Mars surface. This product has several advantages. First, it can be used to instantaneously extract atmospheric- and aerosol-corrected near-infrared (NIR) spectra from any location on Mars. Second, several new data maps can be built as discussed here. That includes new global mineral distributions, quantitative mineral abundance distributions and maps of Martian surface chemistry (wt % oxide) detailed in a companion paper (Riu et al., submitted). Here we present the method to derive the global hyperspectral cube from several hundred millions of spectra. Global maps of some mafic minerals are then shown, and compared to previous works.

## 1-Introduction

The Martian crust contains the record of all the processes that shaped it, from initial differentiation and volcanism, to modification by impact, wind, water and ices. Determining the nature and evolution of these geologic processes is one of the key objectives of Martian exploration, and requires noticeably quantitative measurement of mineralogy and chemistry. A few decades of remote sensing and *in situ* exploration provided considerable advances in the understanding of the Martian crust throughout the geologic ages. In particular, orbital remote sensing with high spatial and spectral resolution in the near-infrared, with OMEGA (Observatoire pour le Minéralogie, l'Eau, les Glaces et l'Activité) on board Mars Express (Bibring et al., 2004) and CRISM (Compact Reconnaissance Imaging Spectrometer for Mars) on board MRO (Murchie et al. (2007)), has demonstrated the ability to correlate mineralogy with specific geologic units. Amongst these different orbital investigations, the imaging spectrometer OMEGA has provided numerous insights on the surface of Mars including identification and global mapping of anhydrous and hydrated minerals (Poulet et al. 2007; Ody et al., 2013; Carter at al., 2013).

One of the most advanced outputs anticipated from OMEGA data is a quantitative retrieval of mineral abundances from the modelling of spectra of selected terrains exposed on the surface. So far, such an approach was performed on restricted areas of the surface using a radiative transfer model (Poulet et al. 2009a, 2009b, 2014). The purpose of the $M^3$ project (Modal Mineralogy of Mars) project is thus to provide global distributions of Martian surface materials derived from spectral modelling. This high-level science product will then be openly distributed through planetary databases (e.g. PSUP (Poulet et al., 2018)). Given the specific NIR spectral sensitivity of OMEGA, we have divided the diversity of Mars mineralogy in two main classes hereafter referred as type-A (mafic-bearing ones) and type-B (hydrated deposits).

This work is the first paper of a series that will present the $M^3$ project. This first step consists in the construction of a 3-D global image cube containing atmospheric- and aerosol-corrected NIR spectra over the 1.0 µm to 2.5 µm wavelength range distributed over 32 px/° and +/-60° of latitude with a spatial sampling of 1.85 km/px at the equator. An application of this new product, which will be discussed in this paper, is to build new mineralogical detection maps. In a companion paper, this global cube will be used to model NIR reflectance spectra from radiative transfer model in order to estimate the mineral abundances and particle grain sizes of type-A terrains.

After a short description of the OMEGA dataset used for the M³ project, we present in section 2 the method to construct the OMEGA-based hyperspectral 3-D image cube of the Martian surface from the entire OMEGA near-infrared dataset also commonly referred to as the C channel dataset (Bibring et al. 2004; 2005). We then discuss in detail in section 3 the methods that are implemented to validate and study this new high-level OMEGA product as a function of operational and observational parameters. The first and quick application of this new product is global mappings of key mafic minerals. A few examples of such capability will be shown in section 4 and compared with previous OMEGA igneous minerals detection maps, namely pyroxene and olivine.

## 2. Building the global hyperspectral cube

### 2.1. Dataset

We first take into account all OMEGA reflectance spectra acquired in the near-infrared ranging from 0.97 µm to 2.69 µm (the "C-channel"). This near-infrared (NIR) OMEGA dataset corresponds to a total of ~9000 data cubes acquired over 3.6 Martian years. Those data cubes were acquired in an elliptic orbit providing global coverage of the surface from the equator to the poles. Different acquisition modes were selected at the timing of the operations, depending on the spacecraft altitude (i.e. velocity compared to the surface): mode 16, 32, 34 or 128 corresponding to the pixels track widths of the image-cube. These four distinct modes provide spatial sampling from ~300 m to ~5 km for the entire surface. The raw spatial sampling of OMEGA thus only depends on the acquisition mode but is not a function of latitude. The projection process used here (section 2.2), that aims at merging all OMEGA image-cubes, will at the end provide a spatial sampling that is latitude dependent as it will be explained in the following section.

All previous OMEGA investigations were based on reflectance spectra including aerosol contributions that could significantly affect the spectra. For this work, we applied both an atmospheric contribution and aerosol correction developed by Langevin et al. (2007) and Vincendon et al., (2007) and adapted for the mid-latitude OMEGA dataset in Vincendon et al., (2015). The atmospheric transmission of a given spectrum is obtained by comparing $CO_2$ signature at 2 µm to a transmission aquired at the bottom and at the top of Olympus Mons. For the aerosols correction, the photons trajectory is modelled using a Monte-Carlo type simulation, in order to simulate the photons path from the entry into the atmosphere to the instrument as described in Vincendon et al. (2007). The model is applied to different geometries of observation and optical opacities in order to cover the global range of parameters. Reference tables are then built and used to extract the aerosols contribution as a function of wavelength, optical opacity as well as incidence and emergence angles of the observation. The optical opacity is then estimated based on the measured optical depth by the MERs rovers Spirit and Opportunity (Lemmon et al. (2015)). A series of data rejection processes was additionally applied, resulting in the exclusion of cubes (1) containing instrumental artifacts (spurious pixels that were identified in mode 128 images from orbit 523 to orbit 3283 and affecting 5 columns of pixels for specific wavelengths), (2) acquired in "mode 16" (16 pixels track width), (3) exhibiting surface ice signatures (water ice (1.5 µm absorption band > 1%) and $CO_2$ ice (1.53 µm absorption band > 5%)) and (4) acquired during unfavorable atmospheric conditions, mainly observations acquired during the 2007 global dust storm. These filtering processes and their related detection thresholds used for the global mapping are enumerated and explained in Ody et al., (2012) and in Audouard et al. (2014). The aerosol correction modifies the "high threshold" (maximum accepted value) on the optical depth. This threshold is fixed at 2.7 instead of 2, as in previous investigations leading to an increase of the number of cubes in comparison to Ody et al. (2012). This larger value is possible due to the correction of the aerosol contribution systematically applied to all OMEGA observations in this paper. By contrast, the observations of the high latitudes (>60°) are excluded as the composition of the high latitude terrains of Mars are significantly impacted by modern re-surfacing processes and by the presence of ices (Poulet et al., 2008). The final remaining dataset constitutes a sample of 3642 data cubes that were used for constructing the global hyperspectral cube. For each location of the surface (i.e. pixel of the final cube) the label (orbit number and image index) of the corresponding projected cube(s) are preserved and will be made available.

## 2.2. Projection of the individual OMEGA observations

The remaining 3642 individual OMEGA image-cubes are projected, using an equirectangular projection, on a global grid predefined at 32px/° (which corresponds to a spatial sampling of ~1.85 km/px at the equator and ~0.9 km/px at 60° in latitude). In the following, raw OMEGA pixels refer to the pixels of each cube prior to their projection on the grid. Depending on the acquisition mode (32, 64 or 128) and on the footprint in longitude and latitude of the individual observation to be projected, the raw OMEGA pixels may correspond to more than one pixel of the global map or may also not completely fill the projected final pixel. In order to prevent spatial under-sampling, when a raw pixel intersects a grid pixel without filling it completely, the given pixel of the grid will be entirely fill with the raw pixel, as long as the raw pixel fills at least 10% of the grid pixel. It is thus important to keep in mind here that this method favors the filling of the global grid but with the caveat that the initial footprint on the final pixel is not taken into account. As a result, the spatial resolution is degraded for the 32 mode image-cube especially at latitudes close to the equator where several raw OMEGA pixels may correspond to the same final pixel on the global map and will be averaged in the projection process. For the latitude far from the equator the sampling becomes sub-kilometer in the final image-cube while the OMEGA images for both the 64 and 128 modes (that covers most of the surface, see Figure 4) have spatial resolution > km which leads in this case to an under-sampling of the surface. Finally, raw pixels from different individual observations will sometimes correspond to the same pixel of the global grid, leading to the superposition of several (at most 13, Figure 1) pixels from various observations. In this case, a specific merging is performed as explained in the next section.

### 2.3. Merging of the individual OMEGA observations into a global image-cube

Once all individual observations are independently projected using the method detailed in the previous sub-section, they are combined into one global image-cube of the Martian surface (Figure 1a). The spectral channels between 2.50 and 2.69 μm are not used because of the atmospheric correction in this spectral range that generates reflectance uncertainties. The measurement of the atmospheric transmission mentioned in the previous section, predicts an atmospheric transmission close to 0 in this wavelength range, eventually leading to diverging reflectance values for these spectels. This step leads to a total of 120 OMEGA spectels (within the 128 native OMEGA spectels) used to build the 3D map. The selected sampling of 32 px/° (providing at best kilometer-scale spatial sampling while limiting the number of pixels in the global reflectance cube) defines a final 3-D product of pixel sizes 11520 (longitude) × 120 (spectral channels) × 3840 (latitude).

In the case of two or more OMEGA projected observations overlapping in the global final map, the image-cubes are merged pixel per pixel (section 2.2). During this step, we do not assign any specific weight to each raw pixel in spite of the different spatial coverage of each observation. Conversely, the reflectance data of those overlapping observations at a given position will be averaged with a weight taking into account the aging of the OMEGA responsivity. The detector of the C-channel that covers the NIR part of the spectrum was impacted over the years by cosmic rays leading to a degradation of usable spectels and hence of the spectral information. We decided to apply a weight to each spectrum during the averaging process depending on the number of viable remaining spectels during the acquisition as a means to track the degradation of the spectral information. It is a function of the percentage of viable spectels (number of unviable spectels divided by the total number of spectels), ranging from 1 to 0.75. This weight can be considered as a tracer of the spectral "quality" of each spectrum. The "hot" (intensity found within 0.8 to 0.95 of nominal intensity) or "dead" (intensity below 0.8 of nominal intensity) spectels are corrected before the projection by performing linear interpolation with closest spectels to fill up the spectral gaps. Depending on the application, the interpolated spectels may be used or not. After this merging step, the global hyperspectral cube covers the surface from -60° to +60° in latitude, with a surface coverage > 90%, and with a ~14 nm spectral sampling.

The largest number of overlapping cubes is 9. On average each spectrum results from the merging of 2.6 OMEGA observations as shown on Figure 1 (b). Actually, 71.5 % of the surface is observed with a combination of two or more observations. The processing procedure detailed in section 2.4 will lead to the exclusion of one observation on every terrain that were observed 3 times or more which will reduce the average number of overlapping OMEGA observations on the overall surface to 2.2. We note that the most observed regions correspond to the mid-latitudes (between -30° to 15°). Some localized regions such Nili Fossae, Mawrth Vallis and Valles Marineris were also often targeted due to their very high interest in terms of mineralogy. The evolution of the quality of the averaged spectra in comparison with individual spectra will be discussed in section 3.1.

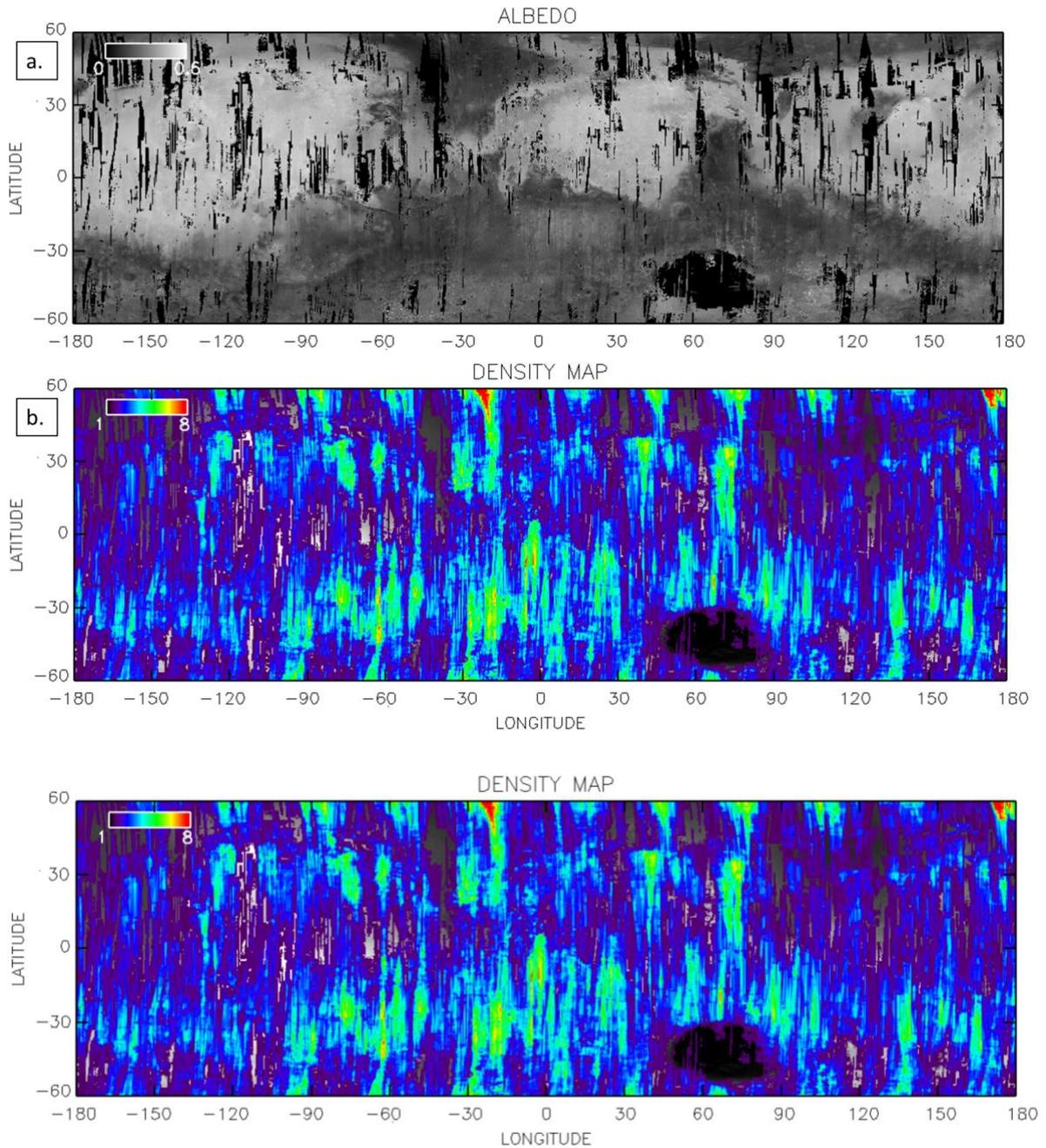

Figure 1 – (a) Albedo map at 1.08 µm obtained from the 3-D global hyperspectral cube. (b) Density map of observations. Color code indicates the number of OMEGA overlapping data used during the merging step, after rejection of unsuitable observations. Red spots correspond to pixels with 8 or more overlapping observations.

2.4 Improvements of the global cube quality based on the variation of albedo

The merging method detailed in the previous section should lead to better quality spectra (section 3.1). However, to provide the most robust product, we decided to perform an additional filtering step based on the albedo standard deviation between overlapping observations. **Error! Reference source not found.**2 (a) presents the standard deviation map of the albedo calculated at 1.08 µm between the different merged observations. The averaged standard deviation (STDDEV hereafter) on all pixels is of 1.9 % of the reflectance at 1.08 µm, including 21.5 % of pixels that are observed only once and whose STDDEV value is set to 0. At first order, this value is rather small compared to the global variation of the albedo (~10 %), which suggests that the filtering and merging processes were well defined and led to a homogeneous map. However, localized areas exhibit large albedo variances. They are mainly found in the north-east part of the Hellas Basin, as well as in some spots of the southern hemisphere. While these variations should not impact the identification of mafic-bearing materials, the quantification of mineral abundances is known to be sensitive to the level of reflectance (Poulet et al., 2009a).

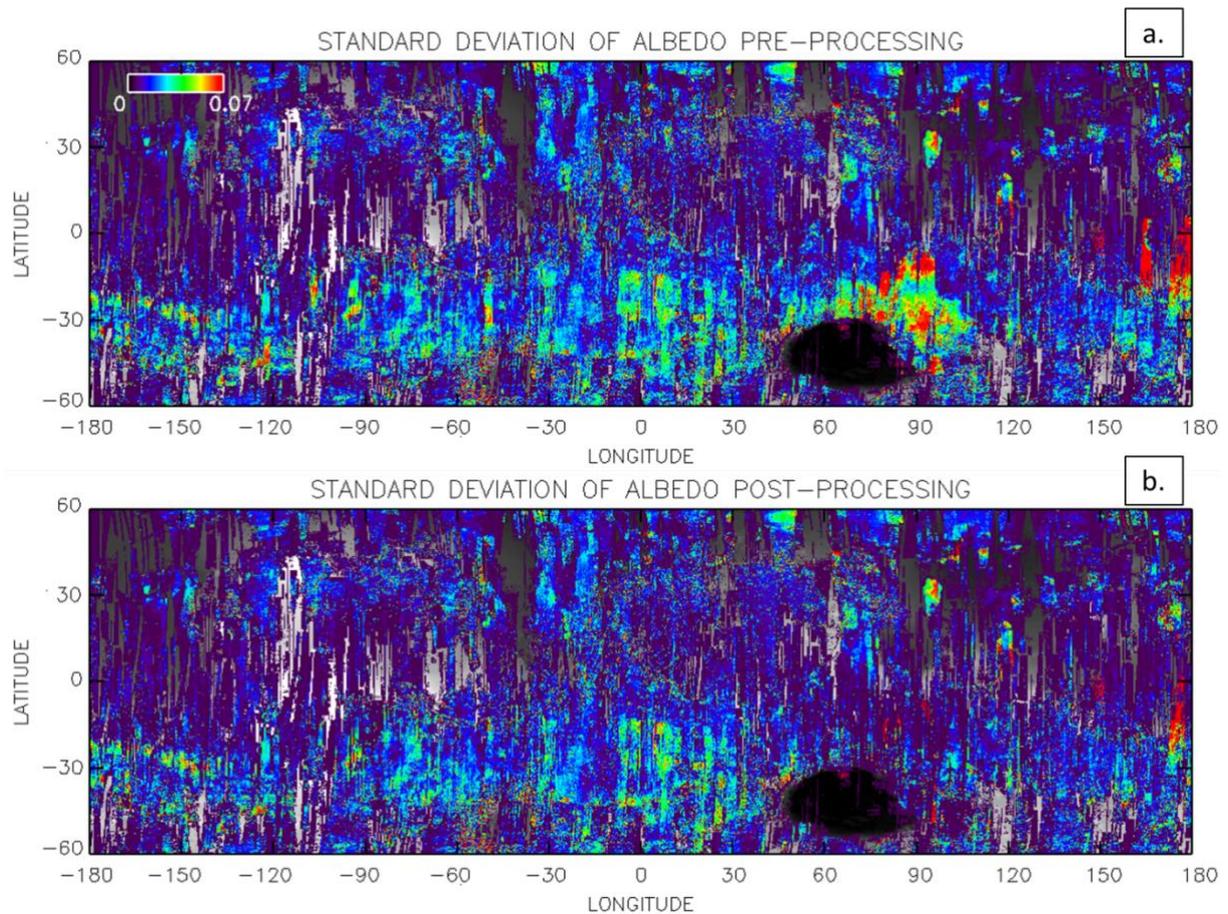

Figure 2 – (a) Standard deviation of the albedo (STDDEV) on the averaged spectra resulting from the merging process. The albedo is calculated with an average value around 1.08 µm. (b) STDDEV albedo map after the additional processing steps described in this section.

Regions with local variations in time and/or in space are easy to identify with the STDDEV map, especially when only one of the different observations at a given pixel can explain the high standard deviation value. We here emphasize few locations where the standard deviation on the albedo is quite suspicious. In the region in the north east of Hellas basin, an increase of the albedo acquired after the major and global dust storm that occurred in 2007 is highlighted (Vincendon et al., 2014). This example confirms that surface changes exist and can significantly affect the averaged reflectance spectra. For this particular region, we only keep the data obtained before the 2007 dust storm. Other localized variations are due to perturbation related to atmospheric phenomenon or instrumental artefact that were not removed by the processing and filtering steps. . For instance, a $CO_2$ icy cloud was spotted in the south of Valles Marineris in one of the observations that was eventually excluded. As the manual inspection remains limited, we decided to apply a systematic search and correction. For each pixel covered by 3 or more observations, we calculate a median albedo, and we systematically exclude the observation that deviates the most from the albedo median value. The number of observations is reduced by one for each pixel without impacting the spatial coverage. A new STDDEV albedo map was then recalculated after this first step of data filtering as represented on Figure 2 (b). We observe that most of the spots that exhibitlarge STDDEV values are removed. This suggests that the high albedo STDDEV prior this additional filtering step is due to the presence of only one outlier image-cube..

### 3. Evaluation of potential biases of the global hyperspectral cube

While the previous global mosaics that resulted from the mapping of individual OMEGA cubes were significantly impacted by sharp spatial variations (Ody et al. 2012), the averaging process used here to construct the hyperspectral cube significantly improves the homogeneity of the mapping. On the other hand, albedo changes are observed on the overall surface as highlighted in section 2.3 (Figure 2). The merged individual OMEGA observations span several Martian years, seasons and local times with three different acquisition modes leading to different spatial resolutions. The data merging does not include any temporal information, either seasonal variations or aging of the instrument (apart from the correction of the dead spectels (section 2.3)) that may impact the quality of the resulting image-cubes as discussed in section 2.3 regarding the spectral degradation of OMEGA data with time. Yet seasonal phenomena (atmospheric variation, dust storm, frost) occurred at shorter temporal scale. Even though spectra are filtered and corrected beforehand, it is worth monitoring the potential impacts of the variations of those parameters on the global cube and whether they can account for the albedo variations. Moreover, the surface is assumed not to be evolving in terms of exposed mineralogy during the 3.6 Martian years of acquisition, in spite of evidence for surface change through dust redistribution at short time scale (Vincendon et al. 2015; Erkeling et al., 2016). It is thus important to keep track whether or not those variations have an impact on the reflectance spectra in terms of Signal to Noise Ratio (SNR hereafter) and albedo spatial variations (Figure 2b). In the following section, several observational and instrumental parameters (acquisition mode, solar longitude $L_S$, dust opacity, incidence angle) are evaluated and correlated with albedo variations to examine potential biases of the global cube as a function of different parameters. Since several parameters may be subject to variations between overlapping observations, it will be difficult to attribute albedo changes to only one parameter. Nonetheless, the observational parameters listed previously can be used as an indicator of the quality of the merging, assuming that small variations of these parameters at a given location will provide better merging quality.

## 3.1. Evaluation of the quality of the averaged spectra

As several observations overlap, the signal to noise ratio of the resulting averaged spectrum should be improved in comparison to any spectrum of individual cubes. The OMEGA spectra of the global cube are on average a combination of 2.2 OMEGA observations (Figure 1b), they supposedly have a higher SNR than individual spectrum. The SNR can only be calculated on the raw DN data before conversion into reflectance data. As a result, to estimate the spectra quality of the highly processed 3D map, we defined a quality criterion (QC) of the averaged reflectance spectra as follows:

$$QC(x,y) = \frac{\overline{REFF(x,n,y)}}{\frac{1}{n} \times \sum_{i=0}^{n} \sqrt{(REFF(x,i,y) - REFF_{smooth}(x,i,y))^2}} \qquad (1)$$

where $\overline{REFF(x, n, y)}$ is the averaged reflectance over $n$ spectels at the $(x, y)$ location, $REFF(x, i, y)$ is the reflectance at the $i^{th}$ spectel at the $(x, y)$ location, and $REFF_{smooth}(x, i, y)$ is the smoothed reflectance at the same pixel obtained by performing spectral window-moving averaging over 3 neighbor spectels centered on the $i^{th}$ spectel. This criterion is used to evaluate the level of high frequency variation (that is considered as noise here) compared to the overall continuum signal level. The computation of this quality criterion (QC) is performed on half of the spectral channels (hence $n = 60$) so as to avoid the spectral ranges where sharp absorption features due to atmospheric or surface contributions could exist (at 1.4, 1.9-2.1 µm and 2.4 µm). With this approach, we estimate the "noise" level, compared to the continuum, considering that the high frequency variations are associated to overall statistical noise rather than actual spectral features.

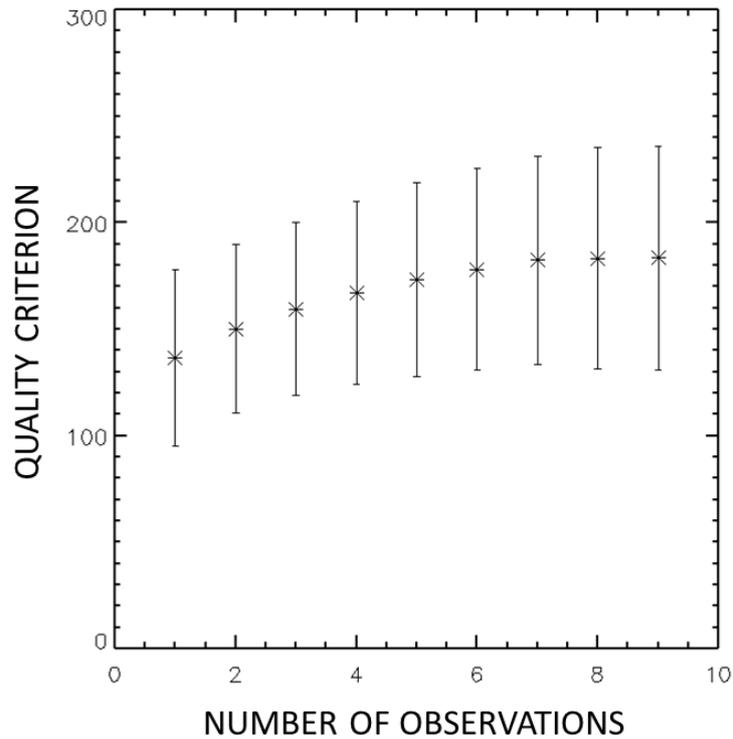

Figure 3 – Evolution of the average quality criterion (QC) with respect to the number of overlapping observations. The error bars represent the quality criterion (QC) dispersion. The calculation was performed by using all pixels.

The evolution of the quality criterion as a function of the number of merged observations is shown on Figure 3. It confirms that the "quality" of the spectra (as defined by eq. 1) increases with this parameter. Therefore, this study shows that the high-level processing method of the OMEGA data presented in section 2 provides better quality spectra, or spectra with reduced high frequency variations, than when individual observations are used.

3.2 Spatial sampling

The individual observations acquired with different acquisition modes (32, 64 or 128 pixel track width) have different spatial sampling and do not have the same footprint on each projected pixel of the final image-cube with a 32 px/° sampling. Figure 4 shows the spatial distribution of track width of the global cube. For the pixels where observations overlap, an average of the swath width is calculated the same way as is the average reflectance, i.e. with a weight assigned to each observation proportional to the number of viable spectels at the time of the observation (see section 2.3).

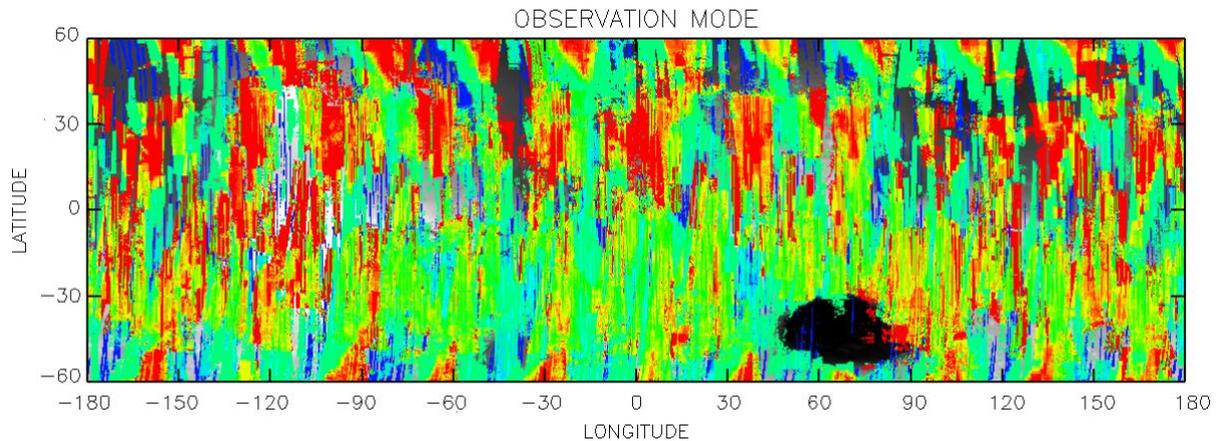

Figure 4 - Global map of averaged track width (correlated to the spatial sampling) of the global hyperspectral image-cube. Red: 128 mode (1.8 to 5.5 km/px). Green: 64 mode (1.1 to 1.8 km/px). Blue: 32 mode (0.55 to 1.1 km/px). The yellow and orange areas correspond to spatial resolution between that of mode 128 and 64. Blue areas correspond to high spatial resolution.

Globally, only 5% of the mapped surface corresponds to the 32 acquisition mode (corresponding to a spatial sampling between 1.1 and 0.55 km/px), 19% corresponds to the 128 acquisition mode (corresponding to a spatial sampling between 1.8 and 5.5 km/px) and finally 25% corresponds to the 64 acquisition mode (corresponding to a spatial sampling between 1.1 and 1.8 km/px). The rest of the surface (51% of pixels) thus corresponds to a combination of several acquisition modes, or spatial samplings. It is important to note that for the 19% of the surface covered only by 128 acquisition mode as well as the locations where the 128 acquisition mode dominates the resulting averaged spectra, one native OMEGA pixel fills more than one gridded pixel thus the actual spatial sampling may be ranging from 1.8 to 5.5 km/px, even if the projected grid provides "better" spatial resolution (1.85 km/px at the equator down to 0.9 km/px at 60° latitude).The merging of observations with different spatial samplings could affect the spectral signature. If a reflectance spectrum of a 32 mode exhibiting a signature is merged with an acquisition of 128 mode in which the signature is not detected due to a mixture at larger scale, it is possible that the spectral signature could no longer be detected in the corresponding pixel of the global cube due to the averaging process. In the following section and in the companion paper, we will nevertheless be focusing on strong absorption features of igneous minerals (identified by pyroxene signature) that are spatially found in extended regions. Their detection and mapping shall thus not be impacted by the merging process. However, for localized and scattered outcrops of olivine and low calcium pyroxene (e.g., Mustard et al. 2005; Poulet et al. 2009c), the averaging process of observations at different spatial sampling for a same pixel could affect the strength of the signature and therefore the resulting modelled abundances. This implies that when studying less spectrally-dominant due to localized exposure phases, the extraction of the individual OMEGA cubes might be necessary to perform spectral modelling of small local deposits. Figure 5 illustrates such an effect on a small olivine-bearing outcrop. While the olivine-bearing deposits are identified using individual OMEGA observations of mode 32 (Figure 5a), they are not detected using the 3D global cube that is the result of 4 merged OMEGA observations with track widths 32, 64 and 128 (Figures 5b and 5c). The olivine was mapped using a spectral parameter from Ody et al. (2012), referred to as OSP1 index. This parameter is used to detect Mg-rich olivine and/or olivine with small grain size and/or low abundance (Ody et al. (2012). The values of the olivine parameter derived for one spot are indicated for each observation. As expected, it strongly decreases with decreasing spatial sampling, and the value derived from the global map is eventually smaller than the detection threshold. This demonstrates that the detection capabilities are limited by the spatial sampling of the native OMEGA observations and thus the spacecraft altitude at the time of measurement (section 2.1). Therefore, because we combined all three observational modes (32, 64 and 128) with the averaging process presented here, the study of small and localized outcrops of minerals may be biased.

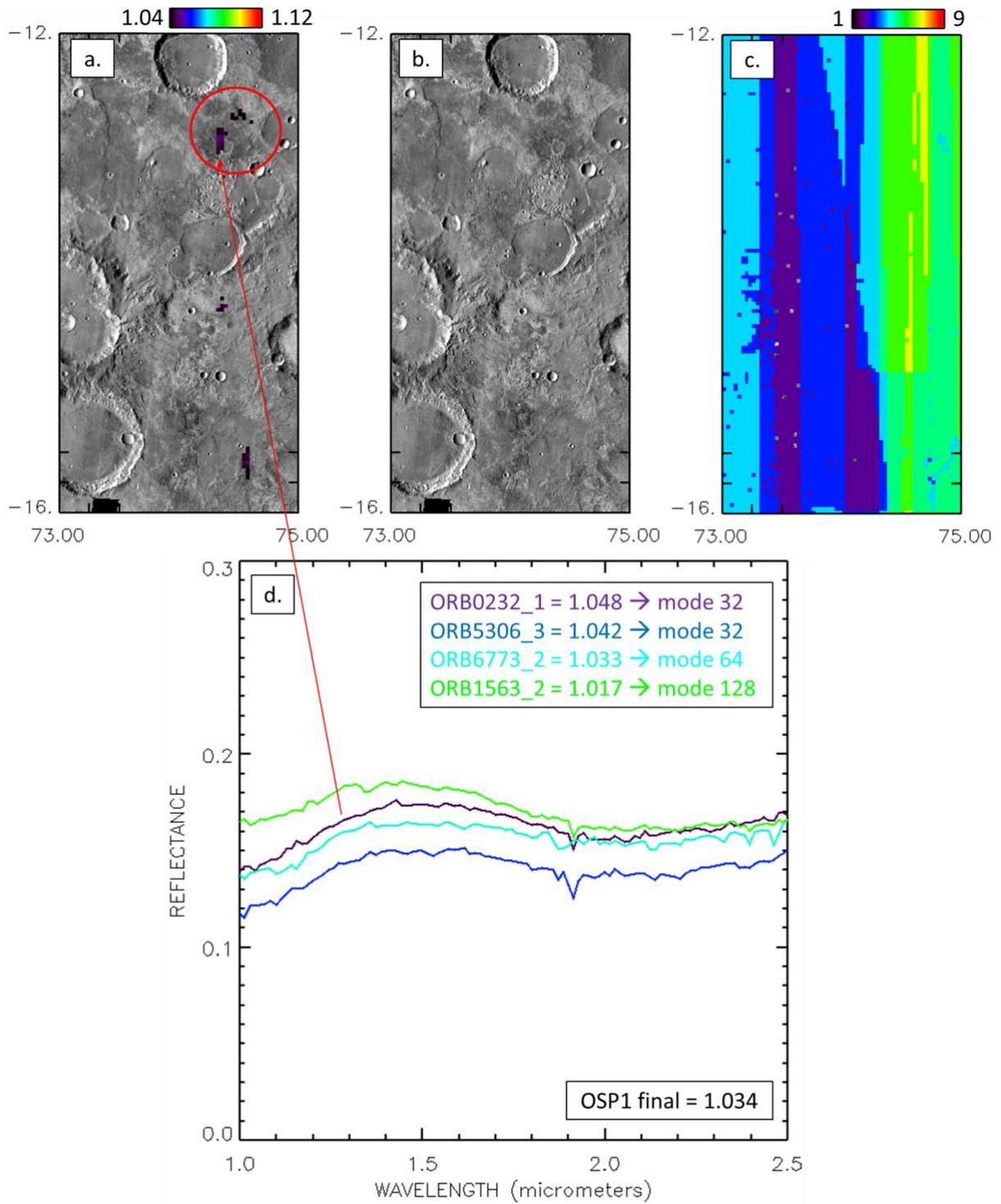

Figure 5 - (a) Individual olivine OSP1 map of observation 0232_1 exhibiting spots of a few pixels with olivine. (b) Olivine OSP1 map derived from the global cube over a region located in the southern hemisphere (Figure 9b). (c) Density observations map (Figure 1b). (d) Corresponding spectra of the different overlapping observations and associated OSP1 values.

### 3.3 Parameters potentially tracing albedo variations

#### 3.3.1 Seasons

The OMEGA individual observations used for the global mapping are acquired during the four seasons of several years, which can be tracked with the solar longitude parameter (Ls). This evolving parameter should enable tracking the possible variability due to the presence of ices, clouds and aerosols, as well as the effects of photometric conditions. Note however that all these effects are taken into account in the processing pipeline, so that their impact on the reflectance spectra of the final cube shall be minimized.

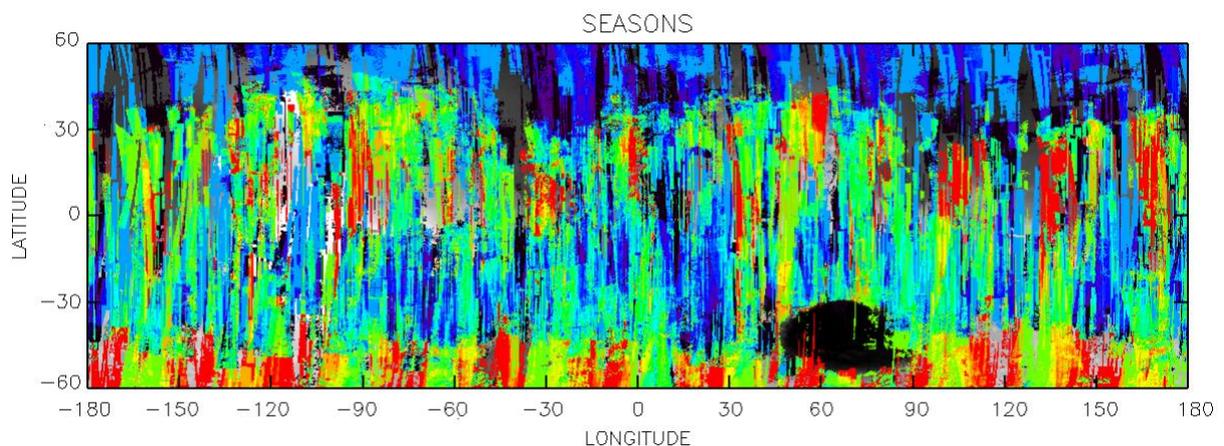

Figure 6 - Global map of the season of observation of the global hyperspectral image-cube.
Blue: spring of the northern hemisphere (0° > Ls ≥ 90°). Green: summer of the northern
hemisphere (90° > Ls ≥ 180°). Yellow: autumn of the northern hemisphere (180° > Ls ≥
270°). Red: winter of the northern hemisphere (270° > Ls ≥ 360°). The green and yellow
regions also account for the combination of different seasons from overlapping observations.

For the regions of latitudes > 30°, the observations were mainly acquired during spring of the northern hemisphere for the northern hemisphere and summer of the southern hemisphere for the southern hemisphere (Figure 6). This is consistent with the filtering processes applied to exclude pixels with ice signatures. These areas that present very minor $L_S$ amplitude shall thus not present spectral variations associated with the seasonal variations. Spectral variations linked to different seasons of observation should be more likely associated to mid-latitude areas where observations acquired during all seasons are overlapping. The standard deviation in terms of $L_S$ is 83° over the entire global cube. The $L_S$ standard deviation for each pixel will be compared to the albedo variation in sub-section 3.3.4 to evaluate if any potential correlation exists.

### 3.3.2 Optical opacity

The aerosols can significantly affect the spectral shape and reflectance level. An aerosols contribution correction is applied for each pixel used for the global map construction. As discussed in Vincendon et al. (2015), the properties of aerosols (optical depth, grain size) are inferred from the typical averaged dust behavior over low and mid-latitudes. As a consequence, we do not account for the exact properties of the atmosphere associated with each OMEGA pixel: the aerosols correction can thus be either under- or over-estimated depending on observations. While the general trend in opacity used for the aerosols correction may not be representative of such small-scaled (in time or space) variations, it provides a good proxy for the overall state of the atmosphere (e.g., higher opacity at lower altitude or during the storm season). As illustrated on Figure 7, the largest optical opacity is found (> 2) in the southern high latitudes, in the vicinity of the impact basins of Argyre and Hellas, as well as in patches located in the mid-northern latitudes (close to 30°N). This trend follows well the climatology of dust optical depth on Mars which shows low optical opacity at equatorial latitudes and an increase when moving towards the high northern and southern latitudes (Montabone et al., 2015).

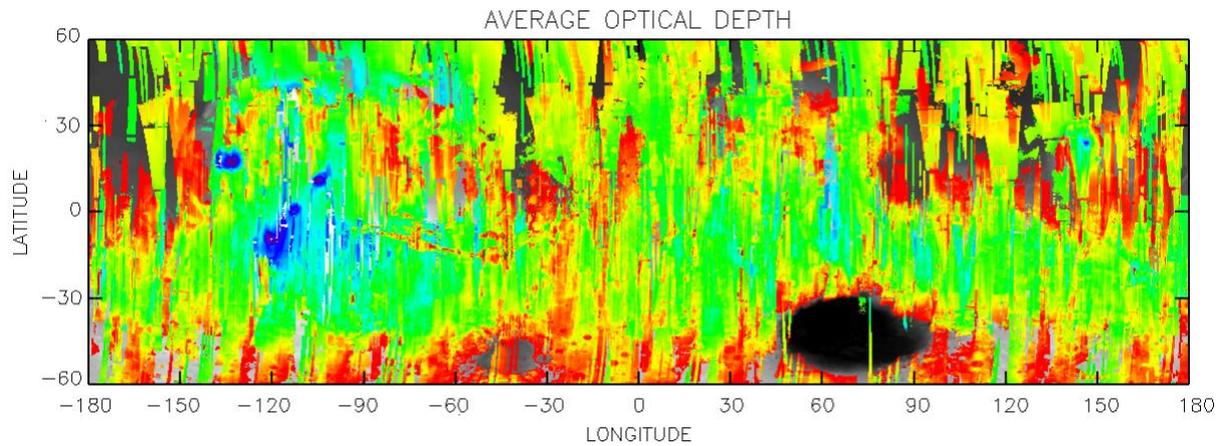

Figure 7 - Global map of averaged optical opacity of the global hyperspectral image-cube from 0 (dark blue) to 2.7 (red). Areas where dust opacity $\tau_{eff} > \mathbf{2.7}$ are excluded.

The average optical opacity is the lowest in the Tharsis region and on its four major volcanic edifices. This observation is of course consistent with the small atmospheric column density at these high altitudes in comparison to the rest of the surface. Although the standard deviation of the optical opacity can reach 1.2 for some pixels, it is on average of 0.24 which is pretty small. As a correction is applied, the impact on the resulting reflectance spectra should be minor. However, as for the seasonal effects, we will assess possible correlations between variations of the dust opacity and albedo variations below, so as to quantify whether this parameter could explain some albedo variations seen in the final product.

### 3.3.3 Incidence angle

Before the merging step, the reflectance of each pixel of every cube is divided by the cosine of the local incidence angle to get the Lambertian reflectance (Audouard et al., 2014). But as for the three parameters discussed previously (track width, season and optical opacity), the values of the incidence angle vary between overlapping observations. It is thus important to monitor the evolution of the incidence angle on the surface depending on the observations. For overlapping observations, we find that the standard deviation among the incidence angles is on average 9° for 90% of the pixels. The majority of overlapping observations hence presents similar incidence angles. We consider that the impact of this parameter on the averaged albedo should be limited. Note that the use of a more evolved photometric function in comparison with the Lambertian case supposed here could be useful to decrease the potential effects of varying geometries of observation on the overlapping spectra (Fernando et al., (2015)).

### 3.3.4 Correlation with albedo variation

Those three last parameters can be used to monitor if any correlation exists with the albedo variations. The variations of three parameters are thus plotted versus the albedo variation (Figure 8). The correlation with the acquisition mode is not relevant because the values being discrete values, the density graph does not add any constraint on a possible correlation with the albedo STDDEV.

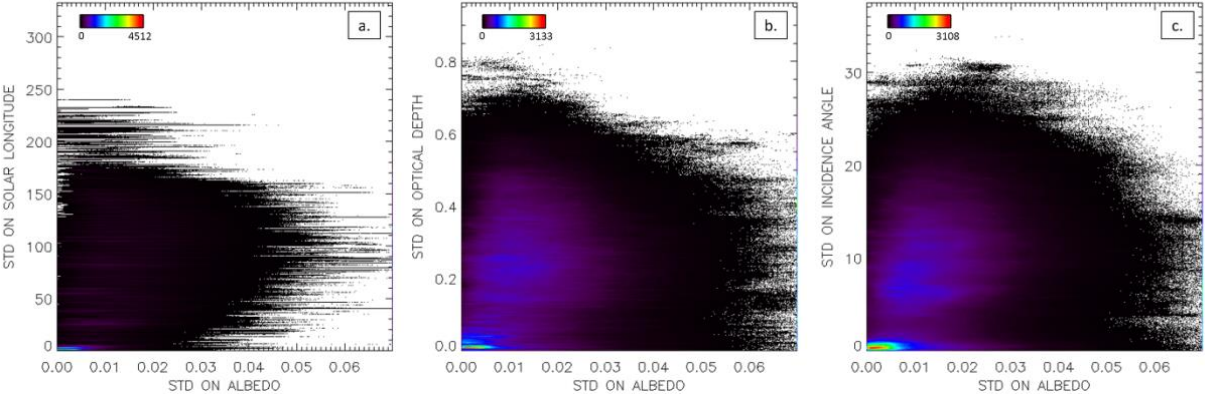

Figure 8 – Density graphs of the standard deviation of three observational parameters with respect to the albedo STDDEV: (a) LS STDDEV. (b) Optical opacity STDDEV. (c) Incidence angle STDDEV.

Although the albedo variation increases with the studied parameters, no major correlation is again detected, and none of these parameters seems to explain the observed large albedo variations (>5%, Figure 2b). This may be due to the fact that for all overlapping pixels, several of these parameters are different, making impossible to isolate one parameter to account for the albedo variations. The high STDDEV, observed on the parameters and on the albedo, can be used as markers of the representativeness of the average spectrum in the 3D cube, and should therefore be monitored in all subsequent regional to local scale studies using this work.

### 3.4 Summary

After all the successive validation, projection and filtering steps described previously, a 3-D reflectance map of Mars was obtained with the following characteristics:

- Spatial coverage/sampling: 85.86% at about $1.85 \times 1.85$ km²/px at the equator and $0.9 \times 0.9$ km²/px at 60° latitudes provided by the projection of each OMEGA image-cube on a grid of 32 px/°. However, depending on the acquisition mode (Figure 4), even if the final grid is based on a fixed spatial sampling of 32 pixel/°, the corresponding spectra can have a native OMEGA spatial resolution ranging from 1.8 up to 5.5 km/px where only the mode 128 is used.

- Spectral coverage/sampling: [0.997 – 2.50 µm] with spectral sampling of 14 nm

- Number of OMEGA observations used: 3642 with 31% of total number of pixels not included

- Total number of raw pixels used for the global image-cube: 91,551,100

This high-level final product will be used to identify (section 4) and to quantify the abundances of the identified mafic minerals. The STDDEV albedo map (Figure 2b) may be used to quantify uncertainties on the resulting spectral indices, modelled abundances and grain sizes.

## 4 -Mapping of mafic minerals

### 4.1 – Detection maps and comparison with previous OMEGA-based works

This study focuses on the identification of mafic minerals and their abundances derived from spectral modelling. The regions to be considered for further analysis are identified using the pyroxene mapping (Figure 9a). The spectral index used to map pyroxene-bearing materials on Figure 9 (a) is the spectral index referred to as "pyroxene" index in Poulet et al. (2007) and Ody et al. (2012). It is based on the 2 μm pyroxene band depth. The individual NIR image-cubes acquired with OMEGA have provided access to identification and global mapping of mafic mineral17(Mustard et al., 2005, Poulet et al., 2007, Ody et al., 2012). Low-calcium pyroxene has been found mostly in Noachian aged units (Mustard et al., 2005; Poulet et al. 2009c), and olivine and high-calcium pyroxene have been found among terrains that cover all geologic units mostly in equatorial to southern low albedo regions (Bibring et al., 2005). The olivine detections have been divided into five distinct types of units and appear more localized than pyroxenes (Ody et al., 2013). The new maps presented here are derived by applying the spectral parameter developed and tested by Poulet et al. (2007) and Ody et al. (2012) to the new 3-D global cube which include the various new filtering and correction methods discussed in section 2 and 3. This results in a larger amount of data compared to previous work. For the study of igneous minerals, we decided to use all interpolated spectels of the 3-D global-cube. Those minerals present broad absorption features that shall not be impacted by this spectral correction.

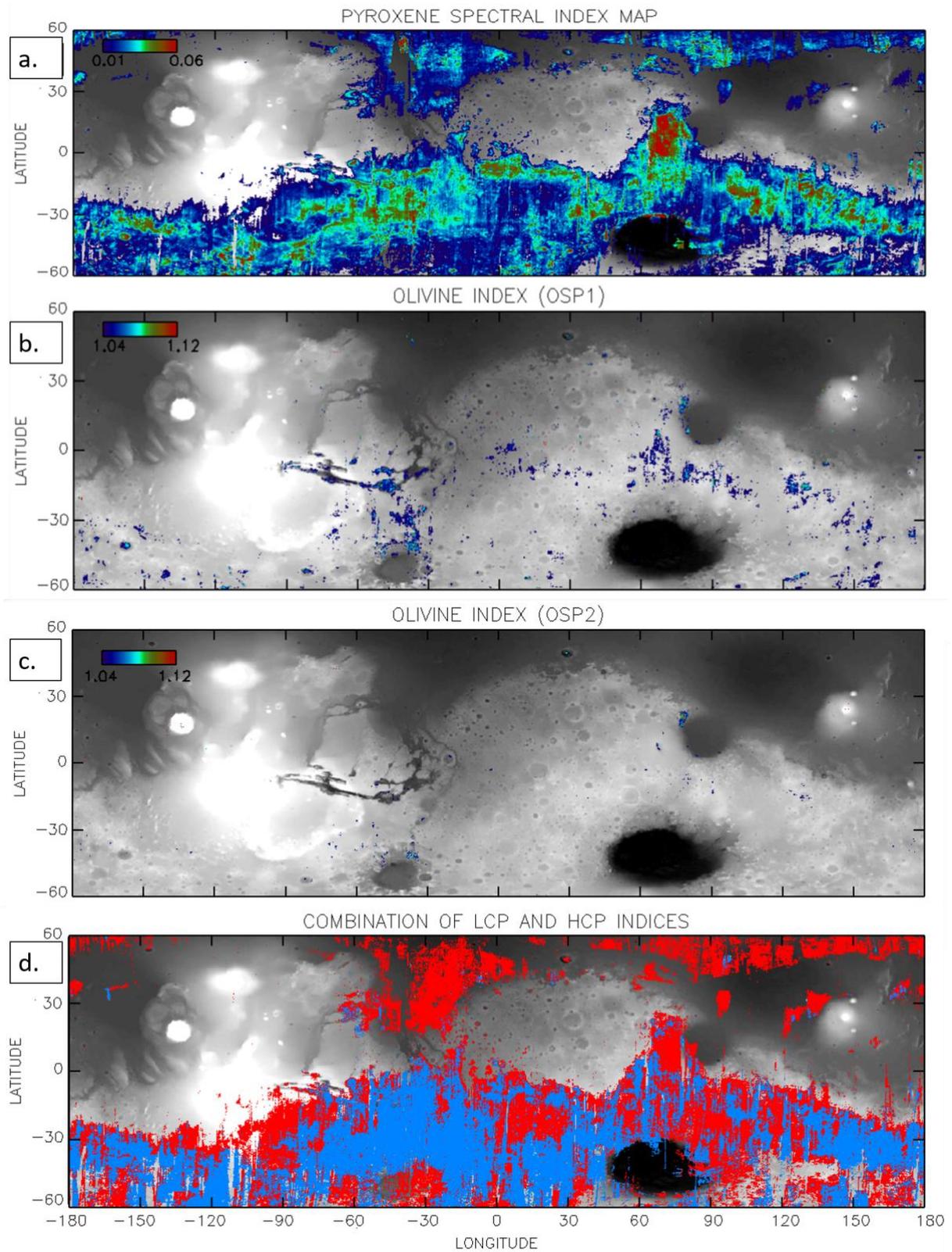

Figure 9 – (a) Pyroxene index map. The detection threshold on the band depth parameter (threshold at 1%) is the same than the one used by Poulet et al. (2007) and Ody et al. (2012). (b) Olivine (OSP1) spectral index map. (c) Olivine (OSP2) spectral index map. (d) Spectral

index map using two spectral indexes indicative of the presence of HCP (HCPINDEX2 of Viviano-Beck et al, (2014) in red), LCP (LCP spectral index of Poulet et al. (2009c) combined with pyroxene index of (a) in yellow) and LCP+HCP (in blue). The absence of widespread yellow terrains is due to the fact that LCP is never detected without HCP. HCP + LCP is shown in blue when HCPINDEX2 > 0.007 and LCP index > 1.065 and pyroxene index > 0.01. HCP is shown in red where HCPINDEX2 > 0.007 and LCP index < 1.065. LCP is displayed in yellow where LCP index > 1.065 and pyroxene index > 0.01 and HCPINDEX2 < 0.007.

The spatial distribution of pyroxene obtained with this method is very similar compared to the previous OMEGA pyroxene maps (Poulet et al. 2007; Ody et al., 2012). We confirm that low albedo regions exhibit strong pyroxene absorption features with dominant HCP, both LCP and HCP and no isolated LCP in the southern hemisphere mostly between 30°S to 60°S in latitude (Fig. 9c). Syrtis Major Edifice presents the largest detection with a spectral index > 0.06 and the rest of the detections in the northern hemisphere are relatively low. This observation is consistent with the TES high-calcium pyroxene map that also predicts enrichment in pyroxene in Syrtis Major and depletion in Acidalia Planitia (Koeppen & Hamilton 2008).

Due to the mapping technique used to compute the 3-D global cube, this map is spatially more homogenous due to the averaging/merging step (Figure 10). This homogeneity is most noticeable at higher resolution as shown on Figure 10. The track-to-track variability of the OMEGA observations is not visible with the new mapping technique (Fig. 10b). Additionally, the detections are well-delimited in the new map and enable a user to highlight correlations with geological units/features that were less spatially-coherent in the previous map (Fig. 10a).

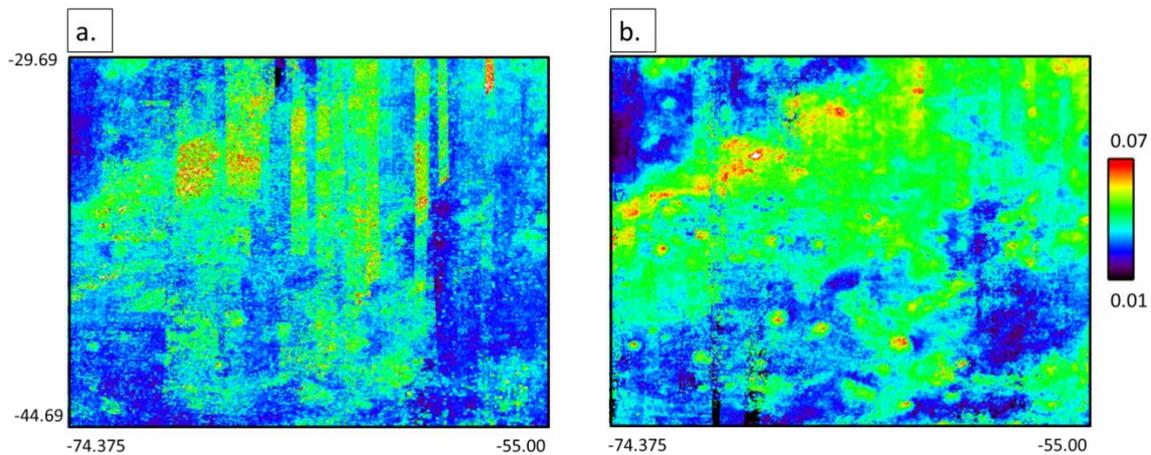

Figure 10 - Local pyroxene detection map. (a) Ody et al. (2012). (b) This study, zoom of Figure 9a.

Two olivine maps are also obtained using OMEGA olivine indices referred to as OSP1 and OSP2 in Ody et al. (2012) indicative of olivine with low iron-content and/or small grain size and/or small abundance (OSP1 - Figure 9b) and high iron-content and/or large grain size and/or large abundance (OSP2 – Figure 9c) respectively. There is almost no olivine detected in the northern plains apart from a few detections in craters floors. Most of the detections are concentrated in the equatorial regions, with larger values in Valles Marineris, south of Syrtis Major and Nili Fossae. The north of Argyre basin is also enriched in olivine (Ody et al., 2013) with many localized detections on its surroundings. The detections for the second type of olivine (OSP2 index) appear more spatially localized and mostly correspond to pyroxene-free terrains (Ody et al. (2013). The largest deposits are restricted to Nili Fossae, north of Argyre and we also observe a few scattered detections in equatorial regions of Valles Marineris and across Terra Tyrrhena and Terra Cimmeria. In other terrains in the southern highlands, type-1 olivine is detected in crater floors and intercrater plains.

The density plots presented in Figure 11 compare the values of the indices for the new detection maps compared to the indices of the latest maps (Ody et al., 2012) for both the pyroxene (a) and type-1 olivine (b) indices. A very good correlation between the two mapping techniques (highest index value from Ody et al. versus index computed from 3D cube) is found. The averaging/merging step globally decreases the spectral index of the pyroxene as expected. For the olivine index the majority of the data points are strongly matching the previous detections. For the low values of the olivine parameter, it is surprising to find values for the average spectra higher than the highest spectral index from individual spectrum. This phenomenon is most likely due to the fact that all spectra are now corrected from the aerosols contribution, which modifies the spectral slope and thus emphasizes the band depths of absorption features (Ody et al., 2012). The good overall agreement between our new pyroxene and olivine maps and previous OMEGA maps can be seen as a first order validation of the 3-D global cube. Additionally, we have also shown here the improvement of the map spatial homogeneousness. This first validation tests coupled with the spectral quality of this new hyperspectral global cube gives us confidence togo a step further and model the pyroxene detections in order to derive the modal composition of the Martian surface as it will be discussed in the companion paper.

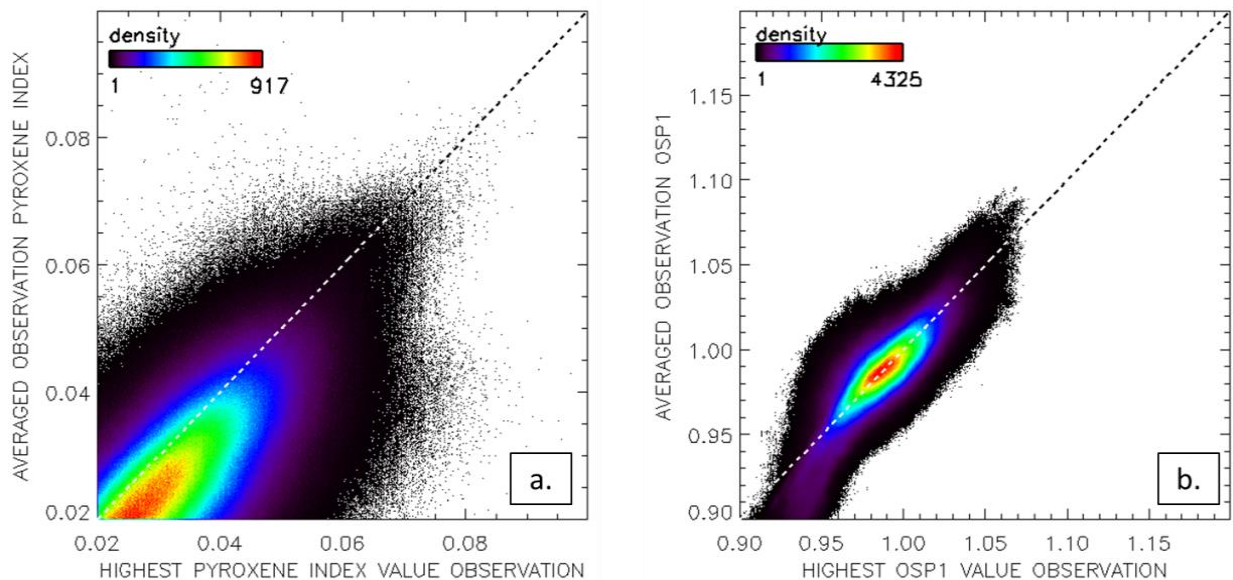

Figure 11 - Density plots comparing spectral index values for the pyroxene (a) and the olivine (b). The x-axis values comes from the Ody et al. (2012) method that consists in selecting the highest index value of any OMEGA cube overlapping a given location. The y-axis corresponds to the value calculated from the 3-D global cube.

Other spectral indices can be easily implemented to compute maps of other mineral phases from this 3-D cube. Moreover the format of the 3-D global map enables an instantaneous visualization of such products and an extraction of spectra of any location of interest. Other mapping techniques such as Modified Gaussian Model (e.g. Clenet et al., 2013) and/or linear spectral unmixing (e.g. Combe et al., 2008) can then also be easily applied. The spectral indices can be built as a way to estimate the band depth (pyroxene and olivine spectral indices of Figure 9 (a) and (b)) or they can be tracers of the overall spectral shape of a mineral (LCP spectral index, Figure 9d), resulting in different detection threshold adapted to each methodology. We present in Figure 9 (d) a combination map of mafic minerals based on OMEGA low-calcium pyroxene spectral index (Poulet et al., 2009c) and CRISM high-calcium pyroxene spectral index (referred to as HCPINDEX2 in Viviano-Beck et al., 2014). The detection threshold for the HCP is the 50-percentile value of the distribution (Murchie et al. (2018)). For the LCP detections the criterion and threshold of 1.05 established by Poulet et al. (2009c) were revisited and adapted to the 3-D global cube mapping. To prevent false detections the detection threshold has been set to 1.065. Additionally, the LCP spectral index that was only based on the shape of the bump around 1.35 µm (Poulet et al. (2009c)), was here combined with the OMEGA pyroxene index (based on the broad pyroxene band depth at ~2 µm) to reinforce the detection accuracy (Figure 9 (a)). These revisited index and threshold enable the detection of spectra consistent with LCP (Figure 12 (b)) and predicts spatial distribution in agreement with the retrived modal mineralogy (Riu et al., submitted). The detection map of pyroxene (Fig. 9d) shows that HCP is widespread in the low albedo regions with or without LCP. Conversely, LCP is detected only in the presence of HCP and is mostly present in the southern hemisphere highlands which explains the absence of yellow terrains on Figure 9 (d). Those spatial distributions are in good agreement with previous investigations including the first based OMEGA analysis (Mustard et al., 2005, Bibring, et al., 2005), but the spatial coverage is here considerably improved. This distribution significantly differs from the TES-based one (Koeppen and Hamilton, 2008). Such discrepancy will be confirmed and discussed in the companion paper (Riu et al., submitted).

Local and regional investigations can be also easily implemented to look for correlation with specific geological units. An example of mineral mapping is shown here over the well-characterized Syrtis Major Edifice. We derive a RGB composite map of the three major mafic phases: HCP, LCP and olivine (Figure 12 (a)). For each unit, we derived average spectrum (over several hundreds of pixels) to highlight the spectral variability of this region (Figure 12 (b)). The surface is dominated by HCP (green on Fig. 12); as previously reported, LCP is observed in more isolated regions especially in the Noachian terrains north of Syrtis Major and is always found in association with HCP (orange on Fig. 12). This mineral is also present on the edges of the edifice. The accuracy of those LCP detections is reinforced by the modelling of abundances which also indicates higher LCP content on the edg (Murchie, et al., 2018)e and north of the edifice (Riu et al., submitted). Strong olivine signatures are found mainly in Nili Fossae (dark blue on Fig. 12) and in more rarely localized patches like impact crater ejecta, but also in combination with both pyroxenes in Nili Patera (light blue and white on Fig. 12). These observations areconsistent with the OMEGA-based analyses previously performed by Clenet et al., (2013) and Combe et al., (2008). We show here that with the global cube we can perform not only global but local analyses of detections and extract a good level of information.

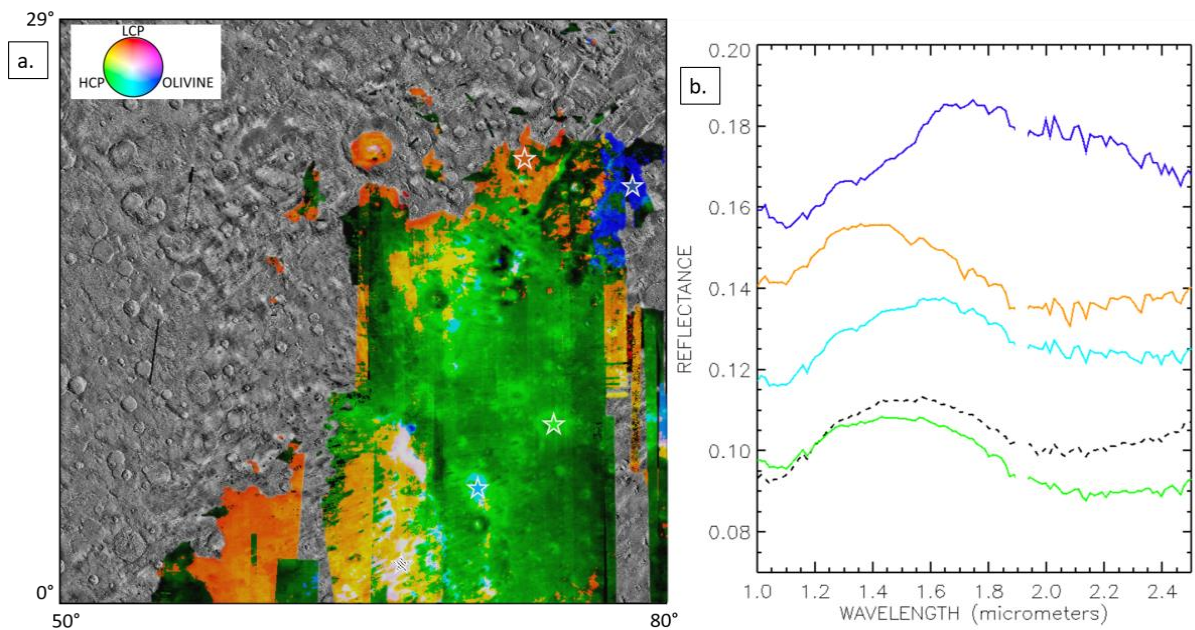

Figure 12 – (a) RGB composite maps of mafic minerals over the Syrtis Major region built with the HCP (green), LCP (red) and olivine (blue) spectral indexes mapped in Figure 9. The white zones correspond to a mixture of the three mineral phases. The darker zones correspond to areas where the different spectral criteria are close to their respective detection threshold. (b) Example of spectra average on several pixels for each type of terrains. Areas used for averaging are indicated by stars on (a). The same color code is used to distinguish the terrains. Green = HCP, dark blue = olivine (mixed with clay signature), light blue = HCP + olivine, orange = LCP + HCP and black dotted line = LCP + HCP + olivine. The hot spectel at 1.9 µm is not corrected

in the global-cube because it does not qualify as a dead spectel and is thus not yet interpolated during processing, we decided to excluded it from the spectra presented on (b).

## 4.2 – Selection of data for the global modelling of mafic minerals

As the major application of the global cube, for the $M^3$ project, is to quantify the mineral abundances at a global scale, we decided to further refine the quality of the global cube for this particular purpose. The objective of the next step is to estimate an albedo STDDEV (Figure 2b) threshold acceptable for the spectral modelling that will be performed on the cube. Figure 13 illustrates the variations of the spectral properties over two locations. While the spectra are rather similar in one case (a) with a small final STDDEV value (0.006), there are strong albedo differences in the other one (Figure 13b) leading to a larger STDDEV value of 0.043. The last case shows large differences in albedo level (likely due to more or less chemically altered "dust" coverage). As the overall albedo is an important parameter to retrieve the surface properties (mineral abundances and grain size), large variations or uncertainties on the absolute albedo value may be propagated into rather large uncertainty on mineral abundances (Riu et al., submitted). A threshold of STDDEV > 0.04 is empirically chosen to exclude pixels from the global mapping presented in further subsequent studies (Riu et al., submitted), leading to the exclusion of about 7% of pixels on the pyroxene detection map that will be used as a starting point for the modelling. This value was chosen as a tradeoff between keeping a good spatial coverage and limiting the potential uncertainties on the modal mineralogy that could come from reflectance variations. For areas that were only sampled once with OMEGA, there is no information on potential reflectance variations with time but they will be used for quantification for spatial covering issues. There will thus be a higher absolute uncertainty on the modal mineralogy of these terrains compared to oversampled terrains with STDDEV < 0.04.

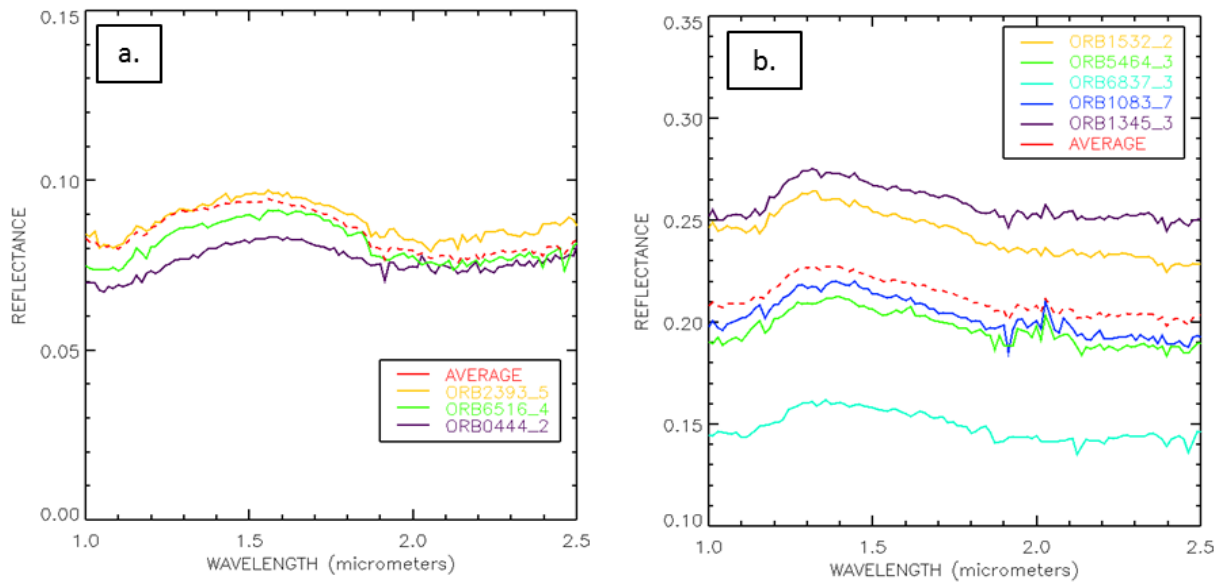

Figure 13 - Example of spectra of same the location with (a) low STDDEV (lon/lat = -107.92°/3.93°) and (b) large STDDEV (lon/lat = 86.99°/-51.08°) compared to the averaged spectrum (dotted red line). A threshold of 0.04 on STDDEV has been selected to remove large variations of albedo.

## 5 – Conclusion

We used the entire OMEGA NIR dataset to build a new highly processed product that will be available to carry out studies of the Martian mineralogy. We describe the technique implemented to combine all observations into a global hyperspectral image-cube ranging from -60° to +60° in latitude with a spatial sampling of 32 px/° over the spectral range 0.997-2.5 µm. This product was compared to several observational parameters of overlapping individual OMEGA observations acquired during several Martian years. The criterion used to evaluate the quality of the resulting averaged spectra is the standard deviation of the albedo level from one observation to the other. We then demonstrated the ability to obtain global maps of key igneous minerals (pyroxenes and olivine). The spatial distributions of the mafic minerals were found to be consistent with previous OMEGA-based global maps. However, the merging step leads to a degradation of the spatial sampling, which potentially precludes the detection of small and isolated spots of olivine. On the other hand, the maps are of higher quality in terms of spatial homogeneity than the previous maps (Ody et al., 2012), which will facilitate the local and regional studies of the correlation of spatial distributions with geological units. Other spectral indices can potentially be applied to the hyperspectral global cube to create new mineral maps. Additionally, application of classification methods and/or deconvolution techniques such as MGM can be envisioned in the future to bring new insights on the distribution of the mineralogy at global scale on the Martian surface and compare in greater detailed the different mapping techniques. Finally, the 3D cube will be delivered to the science community through several web portals including PSUP (psup.ias.u-psud.fr).

Our next objective will be the application of a radiative transfer model to assess the modal mineralogy of the pyroxene-rich terrains. This modelling will be applied to the pyroxene spectral index map (Figure 9a) and only pixels with STDDEV variations smaller than 4% between overlapping observations and with pyroxene spectral index greater than 2% will be kept for the quantification of the abundances. This work will be detailed in the companion paper (Riu et al., submitted) where both the methodology and the results will be discussed. Additionally, this will enable a close comparison with other global studies obtained by TES and GRS (Gamma Ray Spectrometer) as well as with *in situ* measurements.

**Acknowledgements.** This work was partly supported by the French space agency CNES, CNRS and University de Paris Saclay. This project has received funding from the European Union's Horizon 2020 (H2020-COMPET-2015) Research and Innovation Program under grant agreement 687302 (PTAL).